\documentclass[aic]{iosart2x}

\pubyear{2022}
\volume{35}
\firstpage{393}
\lastpage{406}

\begin{document}
\makeatletter
\let\put@numberlines@box\relax
\makeatother

\begin{frontmatter}

\title{Agent-Based Modelling for Urban Analytics: State of the Art and Challenges}
\runtitle{ABM \& Urban Analytics}


\begin{aug}
\author[A,B]{\inits{N.}\fnms{Nick} \snm{Malleson}
\ead[label=e1]{n.s.malleson@leeds.ac.uk}%
\thanks{Corresponding author. \printead{e1}.}}

\author[A,B,C]{\inits{M.}\fnms{Mark} \snm{Birkin}}

\author[B,D]{\inits{D.}\fnms{Daniel} \snm{Birks}}

\author[A]{\inits{J.}\fnms{Jiaqi} \snm{Ge}}

\author[B,E]{\inits{A.}\fnms{Alison} \snm{Heppenstall}}

\author[A,B]{\inits{E.}\fnms{Ed} \snm{Manley}}

\author[A,B]{\inits{J.}\fnms{Josie} \snm{McCulloch}}

\author[F]{\inits{P.}\fnms{Patricia} \snm{Ternes}}


\address[A]{School of Geography, \orgname{University of Leeds},
Leeds, \cny{UK}\printead[presep={\\}]{e1}}
\address[B]{Leeds Institute for Data Analytics, \orgname{University of Leeds},
Leeds, \cny{UK}}
\address[C]{\orgname{Alan Turing Institute},
London, \cny{UK}}
\address[D]{School of Law, \orgname{University of Leeds},
Leeds, \cny{UK}}
\address[E]{School of Social and Political Sciences; MRC/CSO Social and Public Health Sciences Unit, \orgname{University of Glasgow},
Glasgow, \cny{UK}}
\address[F]{Research Computing, \orgname{University of Leeds},
Leeds, \cny{UK}}

\end{aug}

\begin{abstract}

Agent-based modelling (ABM) is a facet of wider Multi-Agent Systems (MAS) research that explores the collective behaviour of individual `agents', and the implications that their behaviour and interactions have for wider systemic behaviour.
The method has been shown to hold considerable value in exploring and understanding human societies, but is still largely confined to use in academia.
This is particularly evident in the field of \textit{Urban Analytics}; one that is characterised by the use of new forms of data in combination with computational approaches to gain insight into urban processes. In Urban Analytics, ABM is gaining popularity as a valuable method for understanding the low-level interactions that ultimately drive cities, but as yet is rarely used by stakeholders (planners, governments, etc.) to address real policy problems.
This paper presents the state-of-the-art in the application of ABM at the interface of MAS and Urban Analytics by a group of ABM researchers who are affiliated with the Urban Analytics programme of the Alan Turing Institute in London (UK). It addresses issues around modelling behaviour, the use of new forms of data, the calibration of models under high uncertainty, real-time modelling, the use of AI techniques, large-scale models, and the implications for modelling policy. The discussion also contextualises current research in wider debates around Data Science, Artificial Intelligence, and MAS more broadly.

\end{abstract}


\end{frontmatter}

\newpage 



\section{Introduction}

Agent-based modelling (ABM) is a subset of the broader field of Multi Agent Systems (MAS), although the terms are often used interchangeably. The characteristics of an agent-based model---such as the use of ``interacting, autonomous `agents'~''~\citep{macal_tutorial_2010} or the general research `mindset' that describes a system ``from the perspective of its constituent units''~\citep{bonabeau_agent_2002}---could be equally applied to MAS. But they differ in that ABM research tends to focus more on the behaviour of living things (humans, animals, cells, etc.) which is not necessarily the case in MAS research. The methodology has proven particularly amenable to modelling human systems with an individual agent often representing a human. The features of the individual (human) agents, including autonomy, heterogeneity, goal-directed behaviour, mobility, reactivity, etc.~\citep{castle_principles_2006, cioffi-revilla_introduction_2014}, allow models to be created that more closely approximate the overall system behaviour than traditional aggregate approaches as they capture the behaviours of the key individual units directly~\citep{schelling_models_1969, epstein_growing_1996, batty_building_2012}.  This is simultaneously the most appealing and challenging aspect of ABM.

One area where ABM is gaining particular interest is in the field of \textit{Urban Analytics}, defined as ``the practice of using new forms of data in combination with computational approaches to gain insight into urban processes''~\citep{singleton_urban_2018}. The use of data and computational approaches to study cities is not new, but in recent years the quantity and variety of data that have become available have created new opportunities for research that requires new techniques and approaches. These new data sources often span multiple spatial and temporal scales---from individual actions such as using public transport or paying for goods, through to macro-level outcomes such as gentrification or population health---and this, coupled with the fact that cities are inherently complex systems that are driven by the actions of micro-units~\citep{batty_cities_2007, batty_building_2012}, puts ABM in a unique position to model urban systems. 

This paper showcases current agent-based modelling research that is being conducted by researchers in the \textit{Urban Analytics} programme (see Section~\ref{sec:turing_ua}) of the Alan Turing Institute, the UK's national centre for AI and data science. It charts the path taken by researchers as they move from small, abstract models of social systems through to all-encompassing, large-scale models that border the realm of \textit{digital twins} (see Section~\ref{sec:large-scale}).

The paper begins by describing the `group' more concisely in Section~\ref{sec:turing_ua}, before moving on to discuss:
\begin{itemize}
    \item the development of techniques used to capture human behaviour in software agents (Section~\ref{sec:behaviour}); 
    \item new forms of data that underpin the field Urban Analytics and their impact on research (Section~\ref{sec:data});
    \item fundamental work on the validation of agent-based models of human systems (Section~\ref{sec:validation}); 
    \item efforts to stream real-time data into agent-based models (Section~\ref{sec:da}); 
    \item work that incorporates AI innovations in ABM (Section~\ref{sec:ai}); 
    \item progress towards representative, large-scale models of urban systems (Section~\ref{sec:large-scale}); 
    \item some of the successful collaborations that have led to these models being used in policy, as well as the barriers that are preventing them from being more widely adopted (Section~\ref{sec:policy}). 
\end{itemize}
The paper concludes with a summary of the future outlook and challenges (Section~\ref{sec:outlook}).

Before continuing, one further definition is necessary. The technique of `microsimulation' is similar to ABM in its focus on individual units (`agents'), yet in a microsimulation model the agents are less likely to interact and may have less nuanced behaviour. There are also differences in the research questions that they are typically used to answer.  Microsimulation is often used to generate synthetic populations that can be used to understand the impact of a policy on individuals within a population over time~\citep{spooner_dynamic_2021, wu_synthetic_2022}. These synthetic populations can subsequently be used in `dynamic` microsimulations, or in agent-based models. That said, there is no clear delineation between the two approaches, and, as with `ABM' and `MAS', the terms are sometimes used interchangeably. In the discussion below we include some research that might be better classified as `microsimulation' rather than `ABM', but it is important that these studies are included as the two methodologies are closely linked anyway, and microsimulation research is having a significant impact on the research agenda more generally.

\section{Agent-Based Modelling  and Urban Analytics at the Alan Turing Institute}\label{sec:turing_ua}

The UK government's ambition to implement a ``ten-year plan to make Britain a global AI superpower''~\citep[][p8]{departmentofdigitalculturemediaandsport_national_2021}, envisages collaborations between ``globally recognised institutes such as The Alan Turing Institute and the high-performing universities which are core to research in AI''~\citep[][p16]{departmentofdigitalculturemediaandsport_national_2021}. It is clear that in the eyes of the UK government, that the Turing has quickly become established as the UK’s national institute for data science and artificial intelligence. Opportunities include: a better understanding of the interactions between health inequalities and environmental factors, such as air quality, and how these new understanding can be used to design more targeted and informed interventions~\citep[][p24]{geospatialcommission_unlocking_2020}; how data can be used better understand crime patterns and to guide interventions that reduce crime in a way that is inclusive and unbiased~\citep[][mission 1]{departmentfordigitalculturemediasport_national_2020}; and intelligent mobility solutions that encourage more sustainable and active travel, reducing the need for parking and thereby freeing up land for new homes or green spaces~\citep[][p30]{governmentofficeforscience_future_2019}. In addressing these challenges, the UK has access to world-class assets through its academic institutions, research council investments, smart city solutions and a diverse ecosystem of business innovation. Furthermore, advanced city modelling and work towards `digital twins' (discussed in Section~\ref{sec:large-scale}) can create a platform to capitalise on international opportunities in which the market for intelligent mobility solutions alone is expected to reach \$1.4 trillion by 2030~\citep[][p33]{governmentofficeforscience_future_2019}.

These opportunities will prove the most fruitful in the context of studying cities because they offer the most fertile domains for ``the use of AI to address societal, economic and environmental challenges in the UK''~\citep[][p14]{departmentofdigitalculturemediaandsport_national_2021}. This recognition led to the creation of the Urban Analytics programme at the Turing in 2018 to provide a focus for sophisticated analytics of big data in metropolitan areas for improved understanding and policy making. The missions of the programme are oriented towards the realisation of healthier and more liveable cities, in which technology supports enhanced mobility, sustainability and equality of opportunity. The remainder of this paper outlines work by the agent-based modelling researchers who are members of the Urban Analytics group (hereafter `the group').

\section{Modelling Human Behaviour in Agent-Based Models}\label{sec:behaviour}

One of the most appealing aspects of agent-based modelling (ABM) is its ability to represent human behaviour and, through simulation, understand how these behaviours play out over space and time.  Our group has made several advances within this area.  \citep{heppenstall_genetic_2007} is an early example of how behavioural rules for agents (in this case petrol stations) were devised using a combination of approaches including numerical analysis, genetic algorithms and published literature.  The behavioural rules were simple and driven by individual petrol stations maximising profit.  Whilst this was appropriate for these agents, replicating the behaviour of human agents requires a more sophisticated approach. Members of the group adopted behavioural frameworks, such as the PECS framework, relatively early~\citep{malleson_crime_2010, malleson_using_2013}---see \citep{schluter_framework_2017} and \citep{muller_describing_2013} for a detailed overview of ABM behavioural frameworks---to allow a more nuanced set of behaviours (in this case, the daily behaviours and motivations of burglars) to be captured and successfully simulated. More recent efforts have included the use of approaches such as Reinforcement Learning, where where positive behaviours are learned through a repeated exposure to an environment. For example, \citep{olmez_learning_2022} use reinforcement learning in the context of a a simple predator-prey model to explore the extent to which the approach can be used to create emergent agent behaviours. Similarly, it  may  be  possible  to teach agents how to navigate spaces as if they were human. These ``spatial learning'' agents may both better reflect the actual behaviours of humans and model their behaviour under changing conditions. Only a handful of simulation models have sought to integrate aspects of spatial cognition and bounded learning~\citep{manley_heuristic_2015, manley_exploring_2018}.

Taking a somewhat different approach, work in the group has also explored how ABM can be effectively used to test hypotheses about typically unobserved behaviours such as those associated with crime and offending~\citep{birks_generative_2012, birks_emergent_2014, birks_street_2017}. Given fundamental logistical and ethical challenges associated with studying such behaviours, these models allow researchers unique means to test the veracity of behavioural theories in generating known empirical system outcomes like commonly observed patterns of crime, or provide  counterfactual realities against which potential crime prevention interventions can be effectively compared. 

The growth in data availability is one of the main foundations for Urban Analytics (see Section~\ref{sec:data}). Higher resolution data can contain information about micro-level behaviours (e.g. the details about interactions between individuals) and, as new forms of data have appeared, researchers have been able to move from relying purely on literature and theory towards using data to both create and validate behavioural rule sets. For example the pedestrian model of \citep{crols_quantifying_2019} (discussed in Section~\ref{sec:validation}) used survey data, sensor data and focus groups to refine behaviour. These data are, however, limited by the events that they do not contain; how can we simulate behaviours that we are interested in if they are not present in the data?  This is a further area where ABM might hold promise; we could use an agent-based model to infer the characteristics of an unobserved behaviour by calibrating it to data on known behaviours~\citep[e.g.][]{vu_multiobjective_2020, greig_generating_2021}. This is the aim of the new field of Inverse Generative Social Science~\citep{vu_inverse_2019}.

Overall, the combination of new, high resolution data coupled with advanced approaches for modelling behaviour---whether through the use of behavioural frameworks, reinforcement learning, or others---is allowing researchers to build behaviours into agent-based models that can be scaled up from models of hundreds of agents to thousands or millions. Section~\ref{sec:large-scale} discusses work that moves towards these much larger models.

\section{New Forms of Data}\label{sec:data}

The Urban Analytics field is characterised by the burgeoning availability of novel data sets. These provide opportunities for the creation of new agent-based models (ABMs) in settings that previously would not have been possible, or for more effective calibration and validation to improve their rigour (discussed in more detail in Section~\ref{sec:validation}). For example: mobile telephone records can offer more detailed cross-sectional or longitudinal movement histories~\citep{sevtsuk_does_2010}; smart ticketing could inform behavioural sensitivities to price or service quality in public transport; smart-phone apps could be used to infer routing behaviour~\citep{malleson_characteristics_2018}; social media or financial transactions could give broader insight into the motivation for travel decisions.  The advantages, and difficulties, in accessing such data have been recognised by learned societies of the major global economies in specific response to the COVID-19 pandemic~\citep{nature_wanted_2021}, but continuing obstacles include licensing restrictions and the maintenance of high standards of privacy and data protection. Furthermore, such data may be readily available in major metropolitan areas such as New York or Singapore, but may not exist, or be inaccessible, in provincial or less industrialised regions. 

Mobility data---i.e. data that describe the movements of people---are a particularly pertinent example of the use of `novel' urban analytics data sets that are being used by researchers in our group to inform their ABMs. For example, \citep{manley_framework_2014} use origin-destination trip data provided by Transport for London to create a `hybrid' agent-based model that represents urban traffic flow driven by complex behavioural models and yet balances this with computational capacity to allow the model to be scaled to large areas with large numbers of agents. In related work, \citep{manley_exploring_2018} use similar transport data to explore how drivers' behaviour and cognition, with respect to route choice, impact the spatial extent and volume of traffic flow within a real-world setting. The results suggest that greater attention should be paid to individual-level models of spatial cognition as these play an important role in predicting urban traffic flow. 

The use of mobility data also extends beyond the study of the behaviour of drivers. New pedestrian count (aka `footfall') data~\citep[e.g. see][]{whipp_estimates_2021} can provide insight into the dynamics of cities, and are being used to build agent-based models of wider urban dynamics. For example, the work of~\citep{crols_quantifying_2019} (discussed in Section~\ref{sec:validation}) uses counts of pedestrians to inform the types of agents that are included in their model and to simulate the daily movement dynamics of people in a small town. At a finer spatio-temporal scale, it is also possible to use pedestrian mobility data to simulate the dynamics of crows of people. This is becoming ever more important as global urbanisation trends~\citep{unitednations_68_2018} lead to larger and denser cities. For example,~\citep{ternes_data_2021} use the trajectories of individual people to create agent-based models that could ultimately be used to manage the movements of crowds in busy urban spaces in real-time (discussed in detail in Section~\ref{sec:da}). 

In summary, there are obvious opportunities offered by these new forms of data. They provide the potential for much more rigorous model calibration or validation (Section~\ref{sec:validation}) and will be key to building larger-scale `digital twin' style models that can support policy making (Sections~\ref{sec:large-scale} and~\ref{sec:policy}). However, there are challenges associated with these `big' data sets that have been well known for some time \citep[e.g.][]{kitchin_big_2013} and care must of course be taken to ensure that the natural biases and other issues around representation are taken into account if they are to be used in modelling.

\section{Calibration, Validation and Uncertainty}\label{sec:validation}

Despite ongoing efforts to improve the empirical rigour of ABMs through the use of robust calibration and validation methods~\citep[e.g][]{thiele_facilitating_2014}, these have remained one of the `key challenges' for the discipline for a number of years~\citep{crooks_key_2008, heppenstall_future_2020, manson_methodological_2020, an_challenges_2021}. This is not surprising for two reasons in particular. Firstly, many evaluation approaches require the underlying model to be executed a large number of times---a million runs is not uncommon~\citep[e.g.][]{vandervaart_calibration_2015}---which is problematic for computationally expensive ABMs. Secondly, the complex, multi-scale nature of the systems that ABMs usually study means that validating a single `pattern' at a single spatial/temporal scale (as would be typical with an aggregate model) would not be sufficient in an agent-based model where multiple patterns at different scales interact~\citep{grimm_patternoriented_2005}, as is the case for most urban phenomena. The degree of uncertainty at one scale is an insufficient measure of the overall uncertainty~\citep{grimm_patternoriented_2005}. 

\begin{figure}[ht] 
    \begin{center} 
		\resizebox{0.9\textwidth}{!}{ 
		    \includegraphics{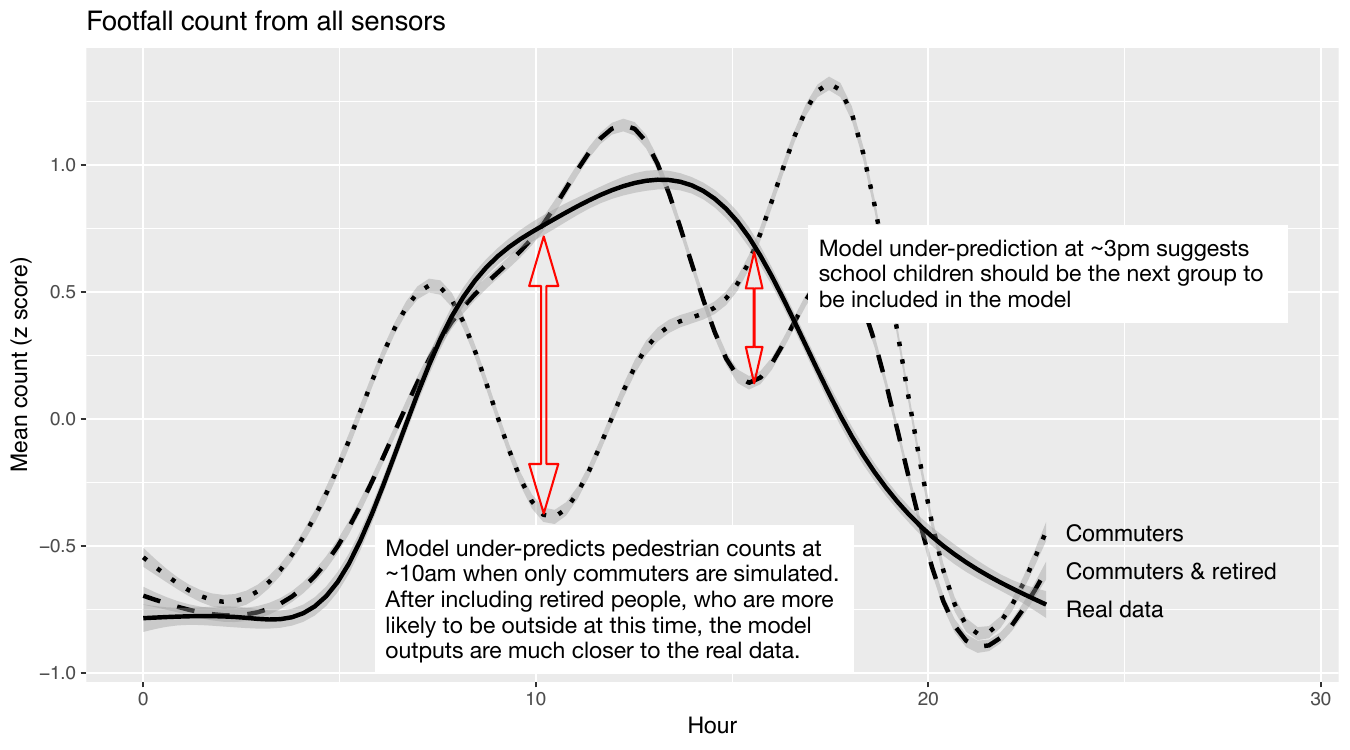}
		} 
		\caption{Comparing average pedestrian counts in two ABMs (one with just commuters, the other with commuters and retired people) to real pedestrian counts recorded using Wi-Fi counters. Adapted from~\citep{crols_quantifying_2019}.} 
		\label{fig:footfall-graph-retired1}
	\end{center} 
\end{figure}

Fortunately the emergence of new urban data sets, as discussed in Section~\ref{sec:data}, offers opportunities for more nuanced calibration and validation of ABMs. As an example, \citep{crols_quantifying_2019} use an agent-based model to study people's daily movements in the town of Otley, UK. In the study area, sensors had been installed to count the number of mobile phones detected in their vicinity over a short time window. Individual phones were not tracked, so the data were anonymous, but the pedestrian counts themselves can give insight into patterns of mobility in the town, showing how many people are in different parts of the town at a high spatio-temporal resolution. \citep{crols_quantifying_2019} created a model that attempted to simulate the daily activities of all individuals in the area, using data from the 2011 census, as they undertook the most common activities (working, shopping, schooling, etc.). The model outputs were compared to the raw data to both calibrate the model parameters but also to determine whether categories of individuals were absent from the model. As illustrated in Figure~\ref{fig:footfall-graph-retired1}, the model significantly under-predicted the pedestrian count when only commuter agents were accounted for. The inclusion of retired people brought the model outcomes closer to the real data, and it was expected that including school age children would also improve the accuracy of the output.

The group are also conducting methodological work that leverages techniques from related fields to support the evaluation of ABMs. Efforts to reduce uncertainty in real-time (discussed in Section~\ref{sec:da}) will help in some situations, but not necessarily for more general calibration/validation challenges. For more general applications, efforts to adapt techniques from the field of Uncertainty Quantification are proving useful. Recent work by~\citep{mcculloch_calibrating_2022} proposes a framework for better understanding the uncertainties associated with an agent-based model and to take these into account when calibrating a model. The authors propose the use of History Matching~\citep{craig_pressure_1997} to rule out \textit{implausible} models and thus reduce the size of the parameter space that needs to be searched during calibration. Calibration is achieved using Approximate Bayesian Computation (ABC) to provide credible intervals over which the given parameters could have created the observed data~\citep{csillery_approximate_2010, turner_tutorial_2012}. The framework therefore allows for the explicit identification and quantification of the various sources of uncertainty in the model, and these are taken into account during calibration or validation.

Related work is also exploring the potential of probabilistic modelling~\citep{ghahramani_probabilistic_2015, davidson-pilon_bayesian_2016}; an approach to model building that provides an elegant framework for representing all forms of uncertainty, updating beliefs about the target system in response to new data, and making predictions.  A \textit{fully-probabilistic} agent-based model, where the agents' variables are represented with probability distributions (i.e. random variables) rather than deterministic variables as is usual, could transform the way that uncertainty is captured and modelled. Initial research has attempted to use bespoke probabilistic programming libraries to simulate a crowd of people in a toy model~\citep{archer_probabilistic_2019} and novel attempts are also underway to develop probabilistic agent-based frameworks that allow for data assimilation~\citep{tang_data_2019} and the calculation of the maximum a-posteriori probability (the mode of a posterior distribution)~\citep{tang_finding_2020}; all of which are useful advances in the development of probabilistic ABM. 

With respect to computational efficiency, often there is little that can be done to speed up the runtime of an agent-based model. Ultimately the models require large numbers of agents to make decisions and to interact, and these processes are computationally expensive. That said, recent work by the group involved re-implementing a large-scale COVID microsimulation~\citep{spooner_dynamic_2021} to reduce its computational complexity using distributed computation technologies such as OpenGL. The gains were astonishing; the time required to execute the model on a desktop PC went from 2+ hours in the original python implementation to under 5 seconds. However, it is worth noting that this was possible because the interactions between individuals were limited, so the improvements are unlikely to be so dramatic when applied to a more traditional agent-based model. In these cases, work on emulators of ABMs might prove useful; for example see~\citep{fadikar_calibrating_2018} who adapt a Quantile Gaussian Process to emulate their model of disease spread. Ongoing, as yet unpublished, work is exploring the opportunities for emulating agent-based models using existing frameworks developed at the Alan Turing Institute (e.g. see~\citep{bowler_urbananalytics_2022}).

\section{Real-Time ABM}\label{sec:da}

\begin{figure}[ht] 
    \begin{center} 
		\resizebox{0.9\textwidth}{!}{ 
		    \includegraphics{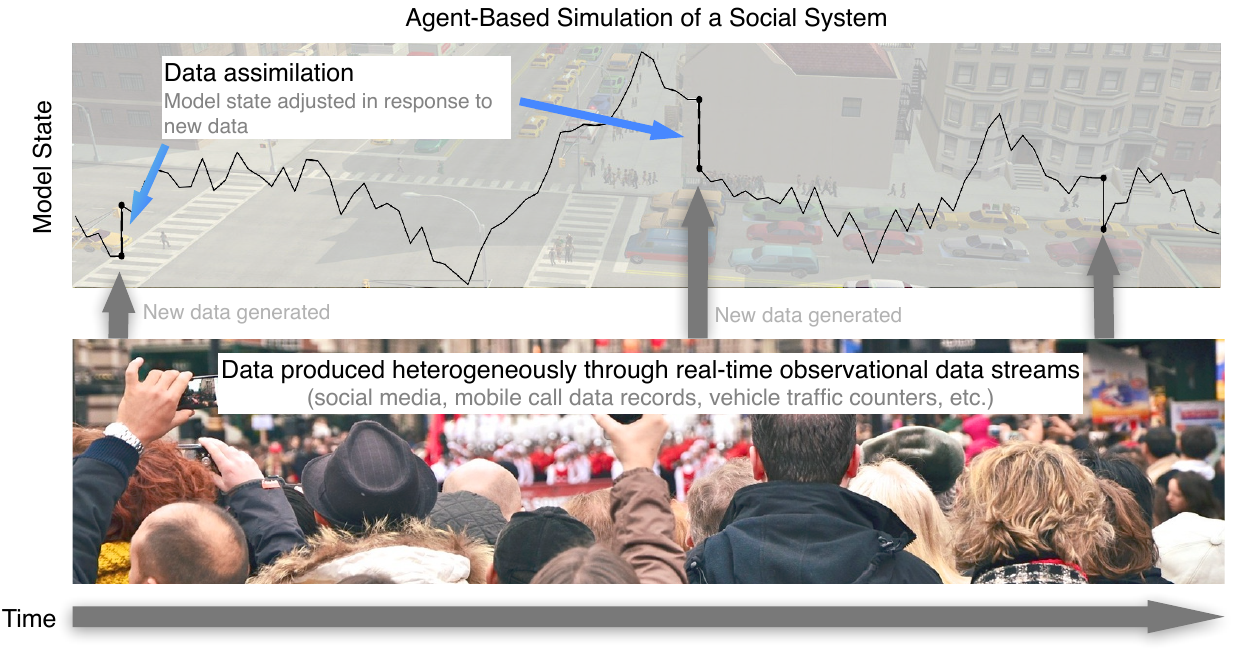} 
		} 
		\caption{An example of a `real-time' agent based model. The model simulates the state of an urban system in real time and draws on up-to-date data to update its state as the urban system evolves.} 
	\end{center} 
	\label{fig:dda}
\end{figure}

Despite important innovations in methods to calibrate and validate ABMs (see Section~\ref{sec:validation}), there remains a fundamental methodological barrier that agent-based modelling has yet to overcome; there are no established mechanisms for incorporating data into ABMs in \textit{real time}~\citep{lloyd_exploring_2016, wang_data_2015, ward_dynamic_2016}. Models are typically calibrated using historical data and then used to make predictions. This is entirely appropriate in many circumstances, such as where systems are in equilibrium, but there are cases where the calibrated model and the underlying system will diverge as they evolve independently. The dynamic re-calibration of an agent-based model's parameter values in response to new data is not uncommon~\citep[e.g. see][]{oloo_adaptive_2017, oloo_predicting_2018}, but adjusting parameter values does not help if a running model has diverged in real-time. In those cases the model \textit{state} needs updating. Fortunately, a body of work has developed in fields that require the integration of real-time data to make more accurate short-term predictions, such as in meteorology~\citep{kalnay_atmospheric_2003}, under the banner of `data assimilation'. Broadly, data assimilation is starting to be recognised as an important barrier and opportunity for ABMs~\citep{heppenstall_future_2020, an_challenges_2021}, and some effort in the group has been devoted to solving data assimilation challenges for ABM. Ultimately, if successful, data assimilation will allow ABMs to make better use of many of the `novel' data sets that are becoming available, adapting in real-time as the social systems that they are tasked with simulating evolve (see Figure~\ref{fig:dda}).

Inspired by efforts that demonstrated data assimilation is possible for relatively simple ABMs~\citep{ward_dynamic_2016}, the majority of the work has focused on the technical challenges of adapting existing data assimilation methods in the context of simulating crowds of people. Crowding was chosen as a test case because crowds exhibit complex behaviour and are driven by the behaviours of individual people, so are an ideal target for simulation with ABM, and because high quality data are readily available for algorithm testing and validation (e.g.~\citep{zhou_understanding_2012}). Initial work~\citep{malleson_simulating_2020} focused on the use of Particle Filters as these are non-parametric so well suited to systems that have non-linear and non-Gaussian behaviour~\citep{long_spatial_2017}. However, a drawback with the Particle Filter is that the number of individual model instances (`particles') that are required to represent the vast space of possible model trajectories grows exponentially with the size and complexity of the system (i.e. the number of agents and the number of agent-level parameters).
Many thousands or millions of particles may be required to successfully represent even relatively small crowds. Whilst there are improvements that can be made to the vanilla Particle Filter to improve its efficiency, alternative methods such as the Unscented Kalman Filter~\citep{clay_realtime_2020} are also being investigated. In addition, work into an entirely new means of defining agent-based models through the use of operators derived from quantum field theory---which is a field ideally suited to representing inherently uncertain systems---may prove transformational~\citep{tang_data_2019}.


A related challenge that the group are addressing, that came to light as they moved from toy to empirical models ~\citep{ternes_data_2021}, is that data assimilation algorithms cannot, by default, optimise \textit{categorical} variables. This type of variable is extremely common in ABM. Therefore attempts are underway to include categorical variables through innovations such as the Reversible Jump, which may allow multiple filters to try to find the `correct' categorical parameter values~\citep{clay_realtime_2021}, or by resampling only some model variables~\citep{ternes_data_2021}. As Section~\ref{sec:outlook} will discuss, the ultimate aim of these efforts is to allow for real-time agent-based models of dynamic urban systems, drawing on diverse data sources and making accurate short-term predictions to aid policy making.


\section{AI in ABM}\label{sec:ai}

The emergence of novel approaches in AI presents ABM research with new opportunities for modelling agent behaviour (as discussed in Section~\ref{sec:behaviour}). These are particularly pertinent for the study of cities where, as well offering opportunities, AI raises new questions as computational agents mix with humans in creating urban dynamics.

The use of AI approaches in modelling agent decisions has become established over the last decade. Statistical classifiers, such as logistic regression, have been a commonly used approach for simulating decisions~\citep{klabunde2016decision}, while computational classification methods, such as neural networks, have been used in simulating vehicle movements~\citep{chong2011simulation} and segregation dynamics~\citep{jager2019replacing}, among other scenarios. As `deep learning' approaches have risen in popularity, these methods have enabled even more data to be included in modelling agent decisions~\citep{https://doi.org/10.48550/arxiv.1706.06302}. Recent developments on modelling trajectories reflects the potential for these approaches within ABM~\citep{10.5555/3398761.3399013}. While Deep Reinforcement Learning (DRL), in part due to its implementation within the Unity ML-Agents library~\citep{juliani2018unity}, has extended to integration of expansive 3D visual inputs within agent decisions. DRL represent a particularly promising avenue for modelling agent behaviours in unseen scenarios and environments, but there is a risk of intractability in agent behaviour. Regardless of the methodology, the increased availability of data, combined with exponentially increasing computational power, enables a greater array of features to be combined in predicting individual behaviours. Nevertheless, questions remain around the trade-offs between the specificity of the behavioural models, simulation scale, computational resources, and the role of theory~\citep{manley2015heuristic} in recreating and predicting future urban dynamics.

Within the swathe of ABMs of cities, one AI technique that has seen little exploitation is Multi-Agent Reinforcement Learning (MARL). Extending the RL framework to multiple agents enables recreation of mutual and social learning, norm formation, and cooperative behaviours, which have implications across various urban problems. However, the technique suffers from challenges relating to the non-stationarity in other agents and environments, which remains an open research challenge in AI more generally~\citep{https://doi.org/10.48550/arxiv.1906.04737}. This area represents one for future exploration and is an ideal candidate for research at the Alan Turing Institute, due to the interdisciplinary nature of the organisation.

Perhaps the strongest contribution ABM can make in relation to AI (as opposed to vice versa) is around understanding the potential future implications of AI for cities. AI is increasingly intertwined with the city. The roll-out of autonomous vehicles (AVs) on city streets is already underway; autonomous lorries, buses, and even aviation vehicles are anticipated over the coming years. While digital technologies have a large impact on the production of urban dynamics (e.g. commuting patterns, shopping habits) that until recently have seemed stable. There is potential for these technologies to prompt a reconsideration of how we design public spaces, roads~\citep{10.1145/3423335.3428163}, housing and land use, and beyond. These technologies have potential implications for social outcomes too, which must be carefully considered and anticipated. The extent to which these technologies will change society remains an open question, and ABM is an important method for exploring these scenarios~\citep{doi:10.1080/03080188.2016.1257196}. At present, the group is undertaking projects that use ABM to explore the potential impacts of AVs on pedestrian dynamics, on elderly mobility, and uptake by different demographic segments of society. Each of these scenarios have important implications for reinforcing or counteracting social inequalities.

\section{Large Scale ABMs and `Digital Twins'}\label{sec:large-scale}

This paper has outlined various efforts that our group is engaging in---such as improving models of agent behaviour, using novel data, advancing calibration and validation methods, incorporating innovations in AI, etc.---to improve the methodological rigour of ABM and to usefully apply ABMs in policy-relevant contexts. If these innovations are successful then they pave the way for the use of ABM as part of much broader computational tools that can simulate human systems over relatively large spatial areas and time periods. This section will outline some of the ways that the group have been exploring the development of larger ABMs before discussing policy-relevant work specifically in Section~\ref{sec:policy}.

The simulation of transportation systems, in particular, has been shown to be possible at increasingly large scales. Using methods similar to those presented in~\cite{manley_framework_2014}, frameworks such as MATSim\footnote{\url{https://matsim.org/}} enable millions of mobile agents to be constructed and simulated to predict mobility dynamics up to country-level scales. Within our group, the CoSMoNorth project is building a simulation of the north of England, an area covering millions of people, for policy testing at scale. On top of the obvious technical/computational challenges associated with such a large simulation, we also hope to develop methods to allow the model to be regularly updated with relevant data, possibly in real time (i.e. through data assimilation as discussed in Section~\ref{sec:da}).

The group also develops spatial ABMs and microsimulations on long temporal scales (10-50 years). These models allow us to address different types of research questions such as the impact of a socio-economic shocks on an urban area and its neighbourhoods \cite{ge2018oil, ge2021simulating}, or the impact of Brexit and changing international relations on food security of cities \cite{ge2021food, ge_not_2018}. These larger-scale transitions and the mid- to long-term ramifications can be hard to pick up in fine-scale data of everyday activities, and require different types of data such as regional or national demographic and economic data, as well as data on migration and social networks. They may also require the researchers to look at different types of behaviour and decision making. For example, the models may look at decisions to buy houses, change jobs, or make investments for retirement, which requires different and perhaps more complex decision making models than those that are appropriate for everyday decisions. There have also been attempts at integrating ABMs at \textit{different} scales to combine multiple systems such as land-use, residential and transport systems \cite{SAEEDI2018214}, using frameworks such as SimMobility \cite{adnan2016simmobility}. Behaviours that occur at different temporal scales (e.g. daily transport mode choice and annual residential choice) can be integrated via the same agents making the decisions.  

Given the range of distinct efforts towards building large-scale models, the Urban Analytics programme has begun to invest in a common platform that combines the simulation of demographic and socio-economic profiles of individuals and households~\citep{lomax_open_2022} with detailed activity patterns across small geographical areas~\citep{batty_new_2021}. The integration of agent-based models with both microsimulation \citep[e.g.][]{lomax_open_2022} and spatial interaction models \citep[e.g.][]{batty_new_2021} is not entirely novel, having been explored previously e.g. in the context of urban housing markets~\citep{jordan_agentbased_2014}, local labour markets~\citep{ballas_gis_2000}, and retail behaviour~\citep{sturley_evaluating_2018}.  However the Turing is now supporting for the first time the integration of microsimulation, spatial interaction and agent-based models at scale within the framework of a `digital twin' of a city with real policy challenges and mechanisms for intervention~\citep{spooner_dynamic_2021}.  

A major challenge with integrating distinct models into a combined framework is the increasing complexity and potential loss of transparency in the integrated models. Individual-level models are already complex, so there is the danger of creating an ``integronster'' \cite{voinov2013integronsters} when the coupling of different modules leads to internal inconsistency (e.g. inconsistent data frequency) that may be difficult to reconcile or even identify. In addition, there is a consensus in some fields (such as economics) that medium-term predictions are the most challenging because short-term predictions are unlikely to be radically different from the current system state and in the long-term systems reach equilibrium. This will be a particular challenge when we try to integrate short- and long-term models to create an overarching `digital twin'.


\section{Towards ABMs as Policy Tools}\label{sec:policy}

Cities are famously now home to more than two-thirds of population of the world~\citep{unitednations_68_2018}, and as such engines of prosperity, opportunity and social progress.  They also provide a canvas for environmental externalities (e.g. pollution, congestion, noise), deprivation, health inequalities and crime.  Such outcomes---both positive and negative---are the product of complex social and economic interactions. This is one key challenges that modellers must face in supporting policy; models---like the systems they seek to emulate---are `messy' and typically do not produce simple predictions. Predictions are often highly uncertain and it is crucial that these uncertainties are communicated properly~\citep{vanasselt_uncertainty_2002, spiegelhalter_risk_2017}, but this uncertainty in outcomes can be challenging for practitioners. Conversely, by representing individuals directly, ABMs can be very accessible and allow practitioners to inform and critique in ways that some other analytical models do not. Hence the demand for comprehensive and effective city modelling tools is escalating within national and local governments, and commercial organisations (ranging from infrastructure providers to retailers), in the UK and internationally~\citep{birkin_big_2021, shi_urban_2021}. 

In a recent and ongoing project, our group has worked with partners in the Alan Turing Institute, Urban Observatory Network, Connected Places Catapult (now home to the UK Digital Twin Hub) and the Department for Transport to establish a prototype of a digital twin for mobility in UK cities.  Drawing on real-time data feeds from three UK cities (Birmingham, Manchester and Newcastle), this demonstrator has already delivered capabilities for decision-support and policy evaluation relating to AI-based traffic simulation, real-time monitoring and forecasting of black carbon, and configuration of low traffic neighbourhoods (LTNs).  For example, in relation to LTNs, travel demand algorithms can integrate historical commuting data with real-time mobility flows to provide a $360^\circ$ view of movement patterns by purpose which are sensitive to diurnal rhythms with fluctuating regional and seasonal cycles.  Onto this base of travel demand and behaviour, microscopic simulations can then predict and evaluate the impact of traffic reduction measures such as traffic calming, pedestrianisation or access restrictions.  In ongoing and future work, the ambition is to provide widespread access to models for any local authority or government department seeking better intelligence and augmented intelligence for enhanced design of urban mobility, infrastructure and land-use.

Recent research by the group funded under the Turing's AI for Science and Government programme is developing ABMs capable of supporting UK Police in better understanding the dynamics of resourcing associated with responding to and investigating crime. Combining spatial micro-simulations that draw on a combination of open source and protected crime data, realistic scenarios of crime-related police demand in any of England and Wales's 43 police force areas can be generated. An agent-based model of police resource allocation that encodes expert insights from operational police officers will allow police to explore how changes in the nature and distribution of crime may affect police agencies ability to respond, and in turn, evaluate the effectiveness of potential new resourcing strategies. Ultimately, these models seek to improve decision-making and support better understanding of complex resourcing problems faced by police~\citep{laufs_understanding_2021}.

Co-production of modelling systems for augmented decision-making will be another critical success factor for agent-based policy models of the future.  This will require collaboration across disciplines and between academic institutions, but most importantly it will rest on the engagement of organisations from business, the third sector and especially government, with the authority and ambition to exploit next generation technologies.  Further methodological development can be expected to yield improvements in handling uncertainty; propagation of errors through complex or linked systems; assimilation of data which are varied, fast-moving and voluminous; and improved representation and parametrisation of human decision-making.  Advances of this kind---many of which our group are currently working on, as outlined throughout this paper---can only help to boost the appeal of models in a policy context. Crucially, the alignment to a broader programme of innovation in AI and Data Science within the Turing will provide a platform for addressing key challenges for example in controlling uncertainty and the propagation of errors, assimilation of massive data sets and machine learning of agent behaviours which are vital in bridging the remaining gap between algorithmic sophistication and augmentation of policy response to real world problems.

\section{Outlook and Challenges}\label{sec:outlook}

This paper has outlined the historical and ongoing research priorities of ABM researchers in the Urban Analytics group of the Alan Turing Institute. It discussed the challenges associated with modelling human behaviour in software agents; outlined how new forms of data that characterise the field of Urban Analytics more broadly are being used to improve the group's ABMs; presented fundamental work on calibration and validation of ABMs, particularly under high levels of uncertainty; outlined preliminary work towards \textit{real-time} agent based models of social systems; presented progress towards large-scale models; and, finally, discussed efforts to use ABMs as key policy tools. 

The outlook for ABM in this area is undoubtedly bright. Only a decade ago models were largely constrained to simulating relatively narrow scenarios---i.e. over small spatial areas, short time periods, or restricted to a narrow set of behaviours---but recent work is delivering models that can operate in much larger contexts. They are being used to simulate systems as diverse as transport~\citep{manley_framework_2014}, crime~\citep{malleson_crime_2010, birks_generative_2012} and farming~\citep{ge2021food, ge_not_2018}, and work towards `digital twins' at the Turing aspires to include ABMs alongside other modules that are already working together~\citep{spooner_dynamic_2021}. In the long run, efforts to better track uncertainty in ABMs~\citep{mcculloch_calibrating_2022}, coupled with closer collaboration with policy makers through the Turing, will begin to foster the use of ABMs as important policy-making tools. 

Despite this optimism, it is important to recognise that ABM in general faces many of the hurdles that were identified a decade ago~\citep{crooks_key_2008, heppenstall_future_2020, an_challenges_2021} and work must aim to address these. The main challenges, as discussed throughout the paper, include issues of scale (how models that work at different spatio-temporal scales should be aligned), model rigour (including how to understand and communicate uncertainty in models and their outputs), computational complexity (as ever, ABMs continue to be extremely computationally expensive) and considerations around ethics and data privacy as models become closer proxies of the real systems that they are simulating. 

As members of a research group, the authors have promoted awareness and relevant capabilities both as individuals and as a team, but also importantly through collaboration within the Urban Analytics programme.  The Turing is ideally placed to lobby for improved access to data; to promote collaborations between disciplines and exploit channels for engagement in government and business; and to accelerate the development of methods and their translation into practice.  The future for agent-based models to support urban analytics and the associated policy challenges is bright, and as a national programme in AI and data science the potential is vast.

\section*{Acknowledgements}

This work in this project has received funding from the European Research Council (ERC) under the European Union's Horizon 2020 research and innovation programme (grant agreement No. 757455), UKPRP (MR/S037578/2), Medical Research Council (MC/\_UU/00022/5) and Scottish Government Chief Scientist Office (SPHSU20). It was also supported by Wave 1 of The UKRI Strategic Priorities Fund (EPSRC Grant EP/T001569/1), specifically within the Criminal Justice System theme of the Alan Turing Institute's research activities, and the Turing's ASG programme (EP/T001569/1, Strategic Priorities Fund - AI for Science, Engineering, Health and Government).



\nocite{*}
\bibliographystyle{ios1}           
\bibliography{bibliography}        

%

\end{document}